\documentclass[aps,prl,superscriptaddress,floatfix,reprint]{revtex4-1}

\usepackage{graphicx}% Include figure files
\usepackage{bm}% bold math
\usepackage{amssymb}
\usepackage{amsmath}
\usepackage{float}
\usepackage{pstricks} % allow the inclusion of post script drawings
\usepackage{color}
\usepackage{bbold}
\usepackage{xr} % allows cross referencing to figures in the supp info
\usepackage{tikz}
\usepackage{comment}
\usepackage{hyperref}
\usetikzlibrary{matrix}

\newcommand{\bra}[1]{\left\langle #1 \right\vert}
\newcommand{\ket}[1]{\left\vert #1 \right\rangle}

\graphicspath{{figures/}}

\begin{document}

\title{Approximating Vibronic Spectroscopy with Imperfect Quantum Optics}

\author{William~R.~Clements}
%\email{william.clements@physics.ox.ac.uk}
\address{Clarendon Laboratory, Department of Physics, University of Oxford, Oxford OX1 3PU, UK}

\author{Jelmer~J.~Renema}
\address{Clarendon Laboratory, Department of Physics, University of Oxford, Oxford OX1 3PU, UK}

\author{Andreas~Eckstein}
\address{Clarendon Laboratory, Department of Physics, University of Oxford, Oxford OX1 3PU, UK}

\author{Antonio~A.~Valido}
\address{QOLS, Blackett Laboratory, Imperial College London, London, SW7 2AZ, UK}

\author{Adriana~Lita}
\address{National Institute of Standards and Technology, 325 Broadway, Boulder, CO 80305, USA}

\author{Thomas~Gerrits}
\address{National Institute of Standards and Technology, 325 Broadway, Boulder, CO 80305, USA}

\author{Sae~Woo~Nam}
\address{National Institute of Standards and Technology, 325 Broadway, Boulder, CO 80305, USA}

\author{W.~Steven~Kolthammer}
\address{Clarendon Laboratory, Department of Physics, University of Oxford, Oxford OX1 3PU, UK}

\author{Joonsuk~Huh}
\address{Department of Chemistry, Sungkyunkwan University, Suwon 440-746, Korea}

\author{Ian~A.~Walmsley}
\address{Clarendon Laboratory, Department of Physics, University of Oxford, Oxford OX1 3PU, UK}

\date{\today}

\begin{abstract}{We study the impact of experimental imperfections on a recently proposed protocol for performing quantum simulations of vibronic spectroscopy. Specifically, we propose a method for quantifying the impact of these imperfections, optimizing an experiment to account for them, and benchmarking the results against a classical simulation method. We illustrate our findings using a proof of principle experimental simulation of part of the vibronic spectrum of tropolone. Our findings will inform the design of future experiments aiming to simulate the spectra of large molecules beyond the reach of current classical computers.\\ \center{Contribution of NIST, an agency of the U.S. government, not subject to copyright.}}
\end{abstract}

\maketitle

\section{Introduction}

\begin{figure*}[htpb]
\includegraphics[width=17cm,angle=0]{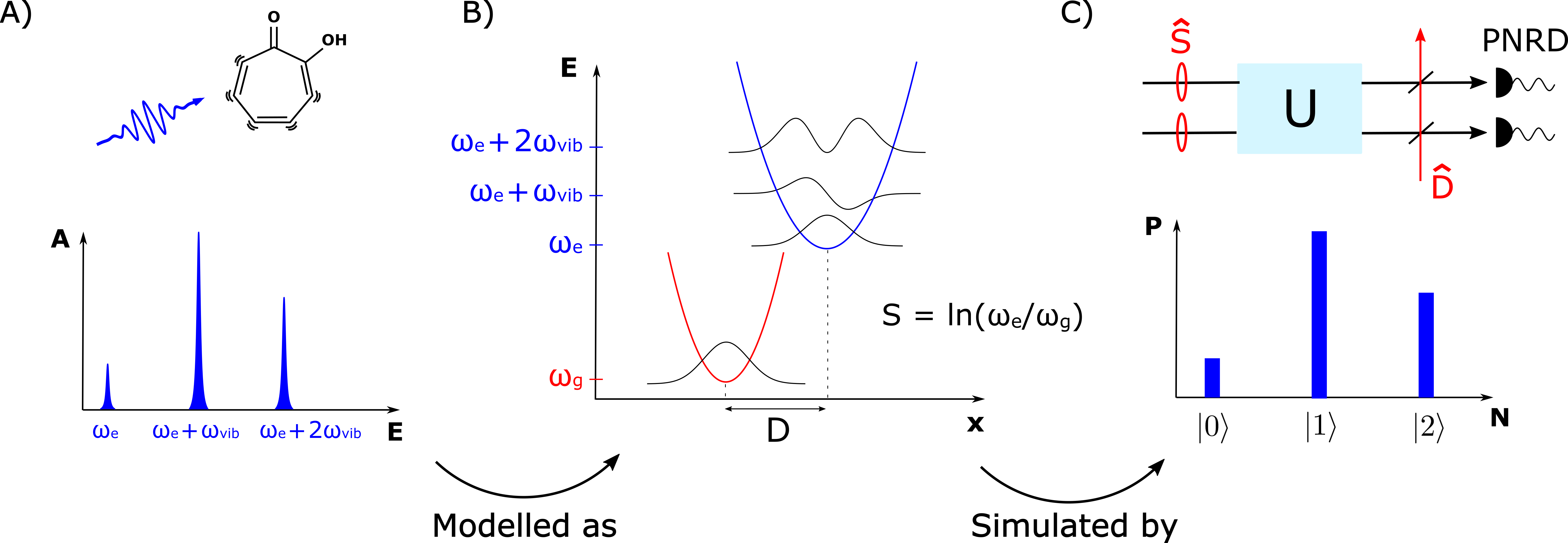}
\caption{Overview of the scheme for estimating vibronic spectra. A) A vibronic transition in a molecule such as tropolone (pictured, top) consists of a joint electronic and vibrational excitation. Depending on the energy of the absorbed photon, different vibrational states are excited, leading to complex spectra (bottom). The heights of the peaks depend on the overlaps between the ground state of the molecule and the excited vibrational states of the excited molecule. B) We model vibronic transitions using a harmonic approximation of the vibrational modes. The harmonic oscillators describing the excited state (in blue) are squeezed by $S$ and displaced by $D$ with respect to the ground state (in red). The overlaps between the different Fock states determine the heights of the spectral peaks. C) We simulate this process using a quantum optics experiment with squeezing $S$ and displacement $D$ (top). Each optical mode is mapped onto a vibrational mode of the molecule; in our case we consider two coupled vibrational modes of tropolone. The probabilities of measuring photon number outcomes using photon number resolving detectors (PNRD) are mapped onto the heights of the peaks in the spectrum (bottom).}
\label{fig1}
\end{figure*}

Quantum chemistry is expected to benefit greatly from the development of quantum simulators and quantum computers, because the ability of classical computers to simulate quantum mechanical processes is severely limited. For example, the calculation of molecular energies can in principle be done on a quantum computer involving relatively few qubits \cite{aspuru2005simulated}.

It has recently been shown that the estimation of molecular vibronic spectra can in principle be done using a quantum optics simulator \cite{huh2015boson,huh2017vibronic}, whereas there is no known efficient classical algorithm for this task. Vibronic spectra, arising from simultaneous electronic and vibrational transitions in molecules, play an important role in determining the optical and chemical properties of those molecules. In addition to being useful for fundamental research in molecular physics and chemistry, calculating these spectra helps in assessing the performance of different molecules for applications in photovoltaics \cite{hachmann2011harvard}, biology \cite{butler2016using}, and other forms of industry \cite{gross2000improving}. 

The protocol for estimating vibronic spectra is considerably simpler than other quantum simulation protocols which require particle interactions or even full quantum computing. Since the vibrational modes of a molecule can be approximated as quantum harmonic oscillators, vibrational transitions can be mapped onto standard operations on quantum harmonic oscillators \cite{janszky1994competition}. An experiment that implements simple transformations such as displacements, squeezing, rotations, and measurements in the Fock basis is sufficient to simulate this physics and therefore reconstruct the vibronic spectrum of a molecule. This protocol scales linearly in the number of vibrational modes to be simulated, does not require any post-selection, and can be implemented with readily available tools in several platforms. It is inspired by the boson sampling protocol \cite{aaronson2011computational} which has been demonstrated experimentally \cite{spring2012boson,carolan2015universal,wang2017high}. 

Various experimental platforms that make use of quantum harmonic oscillators have been used to simulate vibronic spectra, as shown by recent experiments using superconducting devices \cite{hu2017simulating} and trapped ions \cite{shen2017quantum}. Moreover, the original theoretical proposal \cite{huh2015boson} suggests the use of quantum optics, in which each mode of the electromagnetic field is modelled as a quantum harmonic oscillator. The choice of platform imposes practical limitations on what can be achieved. Quantum optics is promising due to the availability of good sources of squeezing \cite{wu1986generation} and the large number of modes which can be manipulated, interfered, and measured \cite{mcmahon2016fully,inagaki2016coherent}. 

However, any experimental implementation of a quantum algorithm on a platform that does not have fault-tolerant architecture is necessarily degraded by imperfections in the system operations. This is a potential limitation to the performance of all specialized quantum processors. In optical platforms, scattering and absorption losses, mode mismatches, detector noise and unanticipated correlations may all contribute to less than ideal operation. The presence of experimental imperfections in any platform was not considered in the original proposal. These imperfections can be expected to affect the simulation, possibly reducing both the accuracy and precision of the results. 

In this work, we explore the impact of these imperfections on a quantum optical simulation of vibronic spectroscopy. We first describe in more detail the analogy between vibronic spectra and quantum optics using a specific transition in tropolone as an example. We then introduce the Gaussian state formalism, which we show can be used to quantify the impact of imperfections. Using this formalism, we propose a method for adapting an experimental setup to account for its imperfections. We also introduce a classicality criterion that can be used to benchmark the performance of an imperfect simulation. Next, we perform a proof of principle experiment in which we simulate part of the vibronic spectrum of tropolone. This experiment highlights the impact of imperfections and illustrates our method for accounting for them. Finally, we discuss our experimental results in light of our analysis.

\section{The 370~nm transition in tropolone}

We first illustrate the connection between vibronic spectrocoscopy and quantum optics using the example of tropolone (C$_7$H$_6$O$_2$), which is a molecule contributing to the taste and color of black tea \cite{giddings1976advances} (see Fig. \ref{fig1}). In the 370~nm electronic transition in tropolone, the change in molecular configuration caused by the electronic excitation distorts the vibrational modes and couples them to each other. In the following, we focus on two of these modes which couple only to each other due to selection rules \cite{smith1998strong}, and use a harmonic approximation of the potential wells corresponding to the vibrational degrees of freedom. We can write the change in mass-weighted normal coordinates $(q_1,q_2)$ of the two modes under study as \cite{duschinsky1937importance, smith1998strong}:
\begin{align}
\begin{pmatrix}
q_1' \\
q_2'
\end{pmatrix}
=
\begin{pmatrix}
    0.9 & 0.436\\
    -0.436 & 0.9
  \end{pmatrix}
  \begin{pmatrix}
q_1 \\
q_2
\end{pmatrix}
\label{Duschinsky}
\end{align}
According to the Franck-Condon principle \cite{condon1928nuclear,franck1926}, the intensity of a given vibrational transition is proportional to the overlap between the wave function of its initial vibrational state and that of its final vibrational state. If the initial vibrational state is the ground state and the final state has $m_1$ energy quanta in mode 1 and $m_2$ energy quanta in mode 2, then this overlap can be written as:
\begin{align}
P(m_1,m_2) = \left|\bra{m_1,m_2}\hat{U}_{Dok}\ket{0,0}\right|^2
\label{P}
\end{align}
\noindent
where $\hat{U}_{Dok}$ is the operator implementing mode transformation \ref{Duschinsky}, known as the Doktorov operator \cite{doktorov1977dynamical}. $P(m_1,m_2)$ is then the normalized intensity of the transition at frequency $m_1\omega_1+m_2\omega_2$, where $\omega_1=176$~cm$^{-1}$ and $\omega_2=110$~cm$^{-1}$ are the excited state vibrational frequencies of modes 1 and 2. $P(m_1,m_2)$ is known as the Franck-Condon factor for this transition. 

We now consider the quantum optics analogy to the vibrational transition described above. Equation \ref{Duschinsky} can be interpreted in quantum optics as a Bogoliubov transformation between two optical modes. If the initial state of these two modes is vacuum, then this transformation can be achieved in quantum optics via two single-mode squeezing operations and a beam splitter \cite{braunstein2005squeezing}. $P(m_1,m_2)$ becomes the probability of detecting $m_1$ photons in mode 1 and $m_2$ photons in mode 2. An ideal quantum optics experiment that prepares two appropriate single mode squeezed vacuums (SMSV), interferes them on a beam splitter with the appropriate reflectivity, and measures the resulting photon number distribution using photon-number resolving detectors can therefore be used to estimate the Franck-Condon factors associated with the 370~nm transition in tropolone. 

We note that in this example, Eq. \ref{Duschinsky} does not include a displacement term. Such a term is present for many molecules. Accounting for this additional term in a quantum optics simulation would require an additional displacement step in the state preparation process.

\section{Theory}

\subsection{Quantifying the Impact of Imperfections}

In the absence of imperfections, equations \ref{Duschinsky} and \ref{P} would be used to determine which squeezing parameters and beam splitter reflectivity to use in an experiment aiming to simulate the spectrum of tropolone. However, in the presence of imperfections these experimental parameters will not yield the targeted photon number statistics. To obtain useful results from an experiment, we require a method for comparing the experimentally generated photon number statistics to those of the target optical state. 

In the limit of a large number of modes, we cannot directly compare the experimentally generated state to the target state in the photon number basis because there is no known efficient classical algorithm for calculating these photon number statistics. We therefore propose to use the Gaussian state formalism \cite{adesso2014continuous,weedbrook2012gaussian} to describe the optical states. This formalism applies to states that have a Gaussian quasi-probability distribution in phase space and to operations that preserve the Gaussian nature of these states, known as Gaussian transformations. Since the initial state (vacuum) and all the optical operations are Gaussian, as well as most realistic sources of imperfection, both the ideal target state and the experimentally generated optical state can be described as multimode Gaussian states. A Gaussian state can be efficiently characterized experimentally in the phase space description, and although the photon number statistics of Gaussian states cannot be efficiently calculated, the fidelity between two such states can be \cite{banchi2015quantum}.

The fidelity between the experimentally generated state and the target state can be used to bound their difference in photon number statistics. The fidelity $F$ between two states described by density matrices $\rho_1$ and $\rho_2$ is related to the trace distance $D$ by \cite{nielsen2010quantum}:
\begin{align}
D(\rho_1,\rho_2) \leq \sqrt{1-F(\rho_1,\rho_2)^2}
\end{align}
$D$ is related to the maximum classical $l1$ distance between different possible measurement outcomes by:
\begin{align}
D(\rho_1,\rho_2) = \max_{\{E_m\}} D(p_m,q_m)
\end{align}
\noindent
where the maximisation is over all sets of detector POVMs $\{E_m\}$ at the output of the network, and $p_m = \text{tr}(\rho_1 E_m)$ and $q_m = \text{tr}(\rho_2 E_m)$. If we consider the POVMs projecting onto photon numbers, we then have that:
\begin{align}
||P_1-P_2|| \leq \sqrt{1-F(\rho_1,\rho_2)^2}
\label{VSinequality}
\end{align}
\noindent
where $P_1$ and $P_2$ correspond to the photon number statistics associated with states $\rho_1$ and $\rho_2$.

Equation \ref{VSinequality} now gives us an efficiently calculable measure for bounding the distance between experimental photon number statistics and those of the ideal state. If $\rho_1$ is the density matrix corresponding to the experimentally generated state and $\rho_2$ is that of the ideal state, we can use this inequality to bound the error on an experimental estimate of the Franck-Condon factors.

In addition to the error bound calculated in this way, we can also account both for the statistical error in estimating the Franck-Condon factors due to the finite number of experimental samples and for small deviations from a Gaussian model of an experiment. If the statistical error is bounded by $\epsilon_\text{stat}$ and the error caused by deviations from a Gaussian model of an experiment is bounded by $\epsilon_\text{G}$, then the distance between the estimated photon number statistics $P_\text{exp}$ derived from the set of measurement results and the Franck-Condon factors $P$ is bounded by:
\begin{align}
||P_\text{exp}-P|| \leq \sqrt{1-F(\rho_\text{exp},\rho_\text{ideal})^2} + \epsilon_\text{stat} + \epsilon_\text{G}
\label{VSgeneralInequality}
\end{align}
\noindent
where $\rho_\text{exp}$ corresponds to the Gaussian description of the experimentally generated state and $\rho_\text{ideal}$ is the density matrix of the target state.

\subsection{Designing an Experiment}

Using the metric described above, we propose a method for determining, in the presence of imperfections, experimental parameters that yield a state that is as close as possible to the target optical state. Typically, an experimental setup has a certain number of experimental parameters that can be controlled, such as the pump power, which determines the squeezing, or the interference between the modes. There also are a certain number of parameters that cannot be controlled such as the loss and the detector dark counts. The task of producing the state with the highest fidelity to the target state can be formulated as an optimisation problem over the controllable parameters given the presence of the uncontrollable parameters. 

To optimize the fidelity, an accurate description of the experimentally generated state must first be formulated as a function of all the experimental parameters. This procedure can be done efficiently since the number of steps required to produce this description is polynomial in the number of modes. Indeed, the squeezers and detectors can all be characterized independently. Efficient methods exist to characterize interferometers \cite{laing2012super}. Losses at the input and output of the interferometer can be individually characterized using classical light. Mode overlaps can be determined using their pair-wise Hong-Ou-Mandel dip visibility. The experimentally generated state can therefore in principle be accurately described for any combination of experimental parameters.

However, the numerical optimization procedure for finding this optimal quantum state is not necessarily efficient. The fidelity can be expected to be a nonlinear function of all the experimental parameters, so the optimization procedure is not straightforward. However, numerical optimization techniques can be used to at least find a local optimum in parameter space, which depending on the desired accuracy of the simulation may be suitable. 

\subsection{Benchmarking against Classical Simulations}

Quantum optical experiments aiming to estimate vibronic spectra are worthwhile if they outperform known classical algorithms. Although the idealised original proposal by Huh \textit{et al} may outperform known classical algorithms, it is not a priori clear that an imperfect experiment designed according to the principles described above would also do so. Furthermore, while there is no known efficient exact classical algorithm for calculating vibronic spectra, some classical approximation strategies do exist. One case-by-case strategy involves guessing which transitions are likely to contribute the most to the spectrum and only calculating the corresponding Franck-Condon factors \cite{jankowiak2007}. A quantum optics experiment with imperfections will only be worthwhile if such other approximation strategies yield worse estimates of vibronic spectra than the experiment.

We propose the following efficient classical approximation algorithm as a benchmark that experiments must outperform in order to produce better than classical estimates of vibronic spectra. This algorithm is conceptually similar to the quantum simulation protocol, so that the same analysis tools can be used in both cases. 

We start by finding the classical optical state, defined as having a regular P-function in phase space \cite{leonhardt1997measuring}, that maximises the fidelity to the target state within the space of all optical states. First, we note that the displacement operation that occurs in the state preparation process can in principle occur before the interferometer instead of after, so that the vibronic spectroscopy experiment consists of squeezed and displaced states sent into an interferometer. The fidelity being invariant under unitary transformations, finding the closest classical state to the ideal target state is equivalent to finding the closest classical state to these initial displaced squeezed states. The closest single mode classical state to a single mode displaced Gaussian state is a coherent state with the same displacement \cite{marian2002quantifying}. Therefore, the closest multimode classical state to the target state is a multimode coherent state with the same displacement. In the case of the transition in tropolone discussed above, the closest classical state is vacuum due to the absence of displacement in equation \ref{Duschinsky}.

Next, we simulate sampling from the photon number statistics of this state. Since this state is classical, its photon number statistics can efficiently be sampled from using a classical algorithm \cite{rahimi2016sufficient}. Since this state is also Gaussian, equation \ref{VSinequality} can be used to estimate the target vibronic spectrum to within some error bound. To outperform this classical algorithm, an experiment must yield an error bound given by equation \ref{VSgeneralInequality} that is smaller than that yielded by the classical state. 

This classical approximation strategy can also be used as a classicality witness for the optical state: any experimental state with a higher fidelity must have a non-regular P-function. We therefore use this best classical state as our benchmark for demonstrating a quantum advantage in experiment. Any experimental optical state that beats the witness is both a non-classical state and produces a better approximation of vibronic spectra than would be possible with any classical state. Furthermore, we note that if an experimentally generated state beats our classicality criterion, then currently known classical simulation algorithms based on the phase space description of the state are generally not applicable \cite{rahimi2015can,rahimi2016sufficient}. 

\section{Experiment}

\begin{figure*}[htbp]
\includegraphics[width=16cm,angle=0]{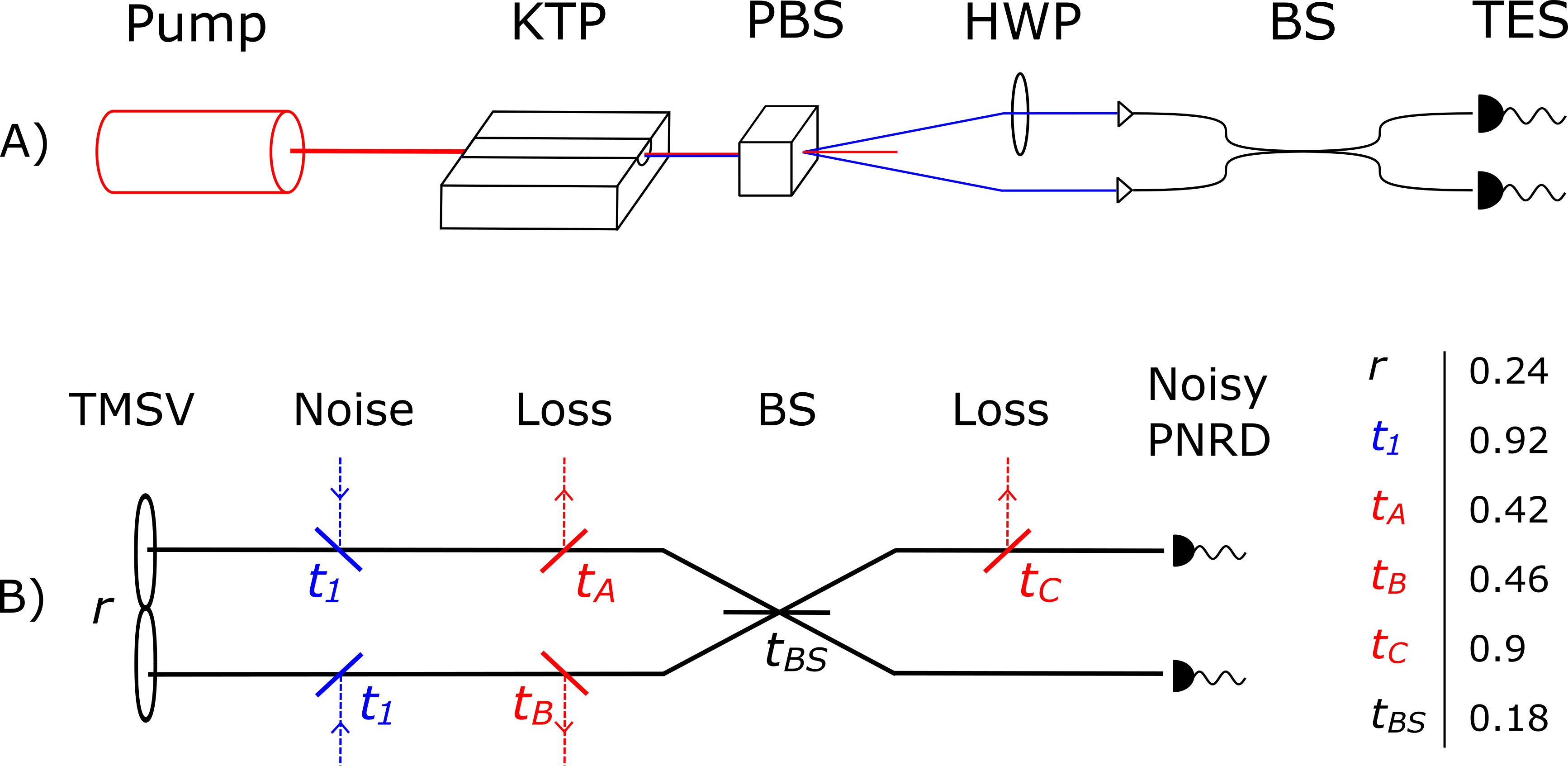}
\caption{A) We approximate the Franck-Condon factors of tropolone experimentally by sending pump pulses at 780~nm into a periodically poled KTP waveguide, separating the two orthogonally polarized downconverted modes at 1560~nm using a polarizing beam splitter (PBS), and rotating the polarization of one of the two modes using a half wave plate (HWP). We then couple these two modes into polarization-maintaining (PM) fiber, interfere them in a tunable PM beam splitter (BS), and measure them using two fiber-coupled transition-edge sensors (TES). B) We model this experiment as a two mode squeezed vacuum (TMSV) interfered on a beam splitter and measured by noisy photon number resolving detectors (PNRD), in the presence of loss and noise produced by the non-overlapping parts of the two modes (both of which are modeled using beam splitters, see Supplementary Materials). The squeezing parameter $r$ of the TMSV and the transmissions of the beam splitters in our model of the experiment are shown in the table on the right.}
\label{fig2}
\end{figure*}

\subsection{Overview}

We perform a proof of principle experiment to simulate the transition in tropolone described by equation \ref{Duschinsky}. We choose tropolone to illustrate our findings for the following reasons. First, we note that the vibronic transition described by equation \ref{Duschinsky} does not include a displacement term, which is present in many other molecules. Displacements can be implemented in quantum optics using classical laser light and do not affect the classicality of a state, as defined by the regularity of its P-function \cite{leonhardt1997measuring}. Furthermore, the squeezing parameters for the two modes in the ideal experiment are $0.19$ and $0.72$, which is quite large compared to most other molecules. The absence of displacement and these large squeezing parameters allow us to focus our analysis on the quantum mechanical aspects of the experiment. These factors also allow us to highlight the impact of imperfections such as loss, which squeezing is strongly affected by, on an experiment.

Our setup (Fig. \ref{fig2}A) consists of a 100~kHz pulsed laser at 780~nm that pumps a periodically poled potassium titanyl phosphate (KTP) waveguide to produce a degenerate two mode squeezed vacuum (TMSV) at 1560~nm \cite{eckstein2011highly}, the two modes of which we separate and couple into the two inputs of a fiber beam splitter. We then use two transition-edge sensors (TES) \cite{lita2008counting} to sample from the photon number statistics from the two outputs of the beam splitter. The frequency of the occurrence of photon numbers $m_1$ and $m_2$ in modes 1 and 2 gives a direct estimate of the joint probability in equation \ref{P} and thus of the Franck Condon factors for the transition towards the states with $m_1$ and $m_2$ vibrational quanta. 

Our setup deviates in the following ways from the ideal setup described above. Firstly, we approximate the two independent SMSVs that are required in an ideal setup using a TMSV, since TMSVs are experimentally simpler to generate and mode-match. The TMSV can be converted into two identical SMSVs using the beam splitter in our setup. Furthermore, the two modes are not exactly identical and so do not interfere perfectly on the beam splitter; the Hong-Ou-Mandel dip \cite{hong1987measurement} between the two modes has a 94\% visibility. Another significant imperfection is the limited system efficiency from the photon source to the detectors; approximately 60\% of the generated photons are not detected. This loss mostly comes from the low coupling efficiency from the photon source into single mode fiber, and is expected to degrade the squeezing. Finally, our TES detectors are noisy, with a 0.2\% dark count probability and a probability of 0.1\% of detecting pump photons that leak through our setup. Since photons from the pump have twice the energy of the downconverted photons and TESs are energy-resolving detectors, these events register as two-photon events.

\subsection{Characterization}

\begin{figure*}[htbp]
\includegraphics[width=17cm,angle=0]{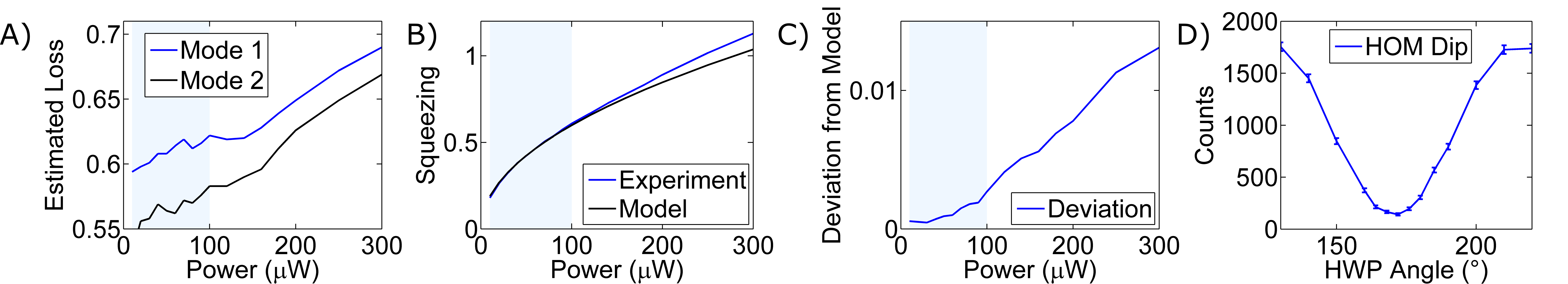}
\caption{Characterization data for our experiment. A) Total losses for the two modes in our setup, estimated from our tomography procedure for different pump powers, with the beam splitter set to full transmission. We use the average and the standard deviation of the values in the shaded area for our estimate of the experimental parameters. B) Squeezing parameter, estimated from our tomography procedure, as a function of pump power. Within the shaded area, we find that the relation between squeezing parameter and pump power is very close to the theoretical relation $P \propto r^2$. C) Deviation between our experimental photon number statistics and the theoretical photon number statistics given by our model for the optical state, quantified by the $l1$ distance. D) Coincidence counts on our detectors measured as a function of the angle of the half wave plate (HWP) in our setup. We observe a clear Hong-Ou-Mandel dip \cite{hong1987measurement}.}
\label{figS1}
\end{figure*}

To produce an optical state that maximizes the fidelity to the target state in the presence of imperfections, we first characterize the experimental setup in order to realise a model for the experiment within the Gaussian state formalism. In the following, we explain how this characterization and modelling is performed. Figure \ref{fig2} shows the model for the experiment that we use as a result of this characterization process.

\subsubsection{Detectors}

We model the TESs as photon number-resolving detectors that, despite having a very low intrinsic dark count rate, suffer from the following noise mechanisms. 

Firstly, our TESs have some degree of inefficiency. This inefficiency is accounted for in our estimate of the total system efficiency described in the following section.

Secondly, the binning procedure that we use to extract photon numbers from the analogue output signal is susceptible to electronic noise on our detectors. With a vacuum input, there still is a 0.2\% probability of wrongly registering a one photon event. We approximate this noise as a dark count mechanism, which can be accounted for within a Gaussian model by considering that our detector is sensitive to both a noiseless input signal and to an additional mode containing a thermal state with a 0.2\% single photon component. 

Thirdly, some pump photons leak through our setup and make it to the detectors. Since TESs resolve the energy of incoming photons, these pump photons are counted as two-photon events. We find that our TESs have a 0.1\% probability of detecting these pump photons. This noise mechanism can be included within our model by considering that our detector is sensitive to an additional pump mode containing a weak coherent state with a 0.1\% single photon component.

This noise can be included in our estimate of the fidelity. Both additional noise modes have a fidelity of about 0.998 to vacuum, and since the fidelity for a product state is the product of the fidelities, the total fidelity of these noise modes to vacuum is 0.9958. This fidelity must then be multiplied by the estimated fidelity of the optical state before the detectors in order to determine the total fidelity.

\subsubsection{Squeezing and loss}

We characterize the total loss (including detector inefficiency) and squeezing in our system using a tomography technique similar to that described in \cite{worsley2009absolute}. We model our photon source as a perfect TMSV with additional loss in the two modes, use our model for the detectors described above, and numerically find the squeezing parameter and the distribution of the loss in both arms which yield the photon number statistics that most closely match the experimental photon number statistics. We follow this procedure for the beam splitter set first to 100:0, and then to 0:100, in order to determine how the losses in the system are distributed in both modes both before and after the beam splitter. We note that since balanced losses can mathematically be commuted through the beam splitter, we need to consider losses in only one of the output ports of the beam splitter in our model.

To verify that our estimate of the losses is reliable, we perform this tomographic procedure for one of the two beam splitter settings for several different pump powers ranging from $10~\mu$W to $300~\mu$W. We find that at powers exceeding 100~$\mu$W our results are skewed as higher order nonlinearities start affecting the pump and the downconverted modes. We therefore use the values for the loss that are found in the low power region in Fig. \ref{figS1}, and estimate the error on these values to be $\pm 2 \%$. These values are also consistent with the heralding efficiencies estimated from the photon number statistics in this plateau region. 

The tomography procedure yields the squeezing parameter that we can directly used in our Gaussian model. The squeezing parameter $r$ is related to the pump power $P$ by the following relation:
\begin{align}
P \propto r ^2
\label{VSpumpr}
\end{align}
\noindent
Once the squeezing parameter has been determined for one power, we can use equation \ref{VSpumpr} to determine the squeezing for any power. To estimate the error on our estimate of the squeezing parameter, we fit the squeezing parameters determined from our optimization procedure to a curve given by equation \ref{VSpumpr} in the plateau region. This fit is shown in Fig. \ref{figS1}. We estimate the error on our estimate of the value of the squeezing parameter at low powers to be $0.01$. 

We also use our tomography results to estimate the error $\epsilon_\text{G}$ stemming from our Gaussian approximation of the experiment. Fig \ref{figS1} shows the deviation, quantified by $l1$ distance, between the measured photon number statistics and those of the closest lossy TMSV measured with noisy detectors determined by our tomography technique. At low powers, this error is less than $10^{-3}$, so we consider that our description of the optical state and of our detectors is satisfactory.

\subsubsection{Distinguishability}

We characterise the distinguishability $\delta$ between the optical modes by using the depth of the Hong-Ou-Mandel interference \cite{hong1987measurement}, shown in Fig. \ref{figS1}, measured by setting the beam splitter to 50:50 and measuring the number of coincidence counts at the outputs as we rotate the HWP in our experimental setup. We use SNSPDs as our photon detectors for this procedure due to their greater ease of operation. Considering that non-overlapping parts of the two modes can be modelled as noise photons, the ratio of noise photons to signal photons in the system is then $\delta$. We choose to model this noise as virtual beam splitters of reflectivity $\delta$, placed just after the squeezing operation for both modes, between each mode and a virtual thermal state containing the same average number of photons as the TMSV. The $\pm 2\%$ error on our estimate of $\delta$ comes from the error on the estimate of the depth of the measured Hom-Ou-Mandel dip. We note that our model for the noise given by the distinguishability is only an approximation of the full description of this noise, which would require taking several additional non-interfering modes into account. However, our model provides a rough estimate of the contribution of this noise towards the degradation of the fidelity.

\subsubsection{Beam splitter reflectivity}

The beam splitter reflectivity was set by blocking one mode in our experiment, setting the beam splitter to be fully transmissive so that the maximum photon number at the detector for that mode could be determined for a given pump power, and then adjusting the beam splitter until the average photon number at the detector was the desired fraction of the maximum. The error on our estimate of the reflectivity was estimated at $\pm$1\%.

\subsection{Optimization and Results}

\subsubsection{Finding the optimal state}

The characterization procedure described previously yields a Gaussian model for the experiment. For any value of the beam splitter reflectivity and squeezing parameter, we now have a Gaussian description of the output state that accounts for the imperfections in the setup. 

To find the values of the squeezing parameter $r$ and beam splitter reflectivity $t_{BS}$ that maximize the fidelity of the experimentally generated state to the target state, we use Matlab's built-in fminsearch procedure. Given the small size of the parameter space, this routine can be expected to find the global optimum for the fidelity. Fig. \ref{fig2}B shows the values of $r$ and $t_{BS}$ that maximize the fidelity within our model. This maximum theoretical fidelity is 0.891.

Given the errors in our model, we expect not to achieve this maximum fidelity in practice. To obtain a more reasonable estimate of the fidelity, we use a Monte Carlo method to determine the most likely value of the fidelity that we achieve as well as the error on this value. We simulate 100 states produced by our experiment, where we randomly select the experimental parameters from a Gaussian probability distribution determined by the estimated mean and standard deviations given by our analysis. The mean and standard deviation in the fidelity of these samples are used as our fidelity estimate and as the error on this estimate. With this method, we revise our estimate of the fidelity to the target state to 0.890(1).

\subsubsection{Estimated Franck-Condon factors}

We collected 1638370 samples over the course of about 20 seconds from the photon number statistics of our state from which we estimate the Franck-Condon factors of the transition under study. Table \ref{ExpResults} compares our experimental estimate to the exact theoretical target values numerically calculated using equations \ref{Duschinsky} and \ref{P}. By comparing our experimental values to the exact values, we find an error of 0.206. This error is quite large; the following section will provide a detailed analysis of the sources of error.

\begin{table}[ht]
\centering
\begin{tabular}{ l c c c r }
  Frequency & \textbf{Experiment} & \textbf{Ideal} \\ \hline
  0 & 0.9628 & 0.7731 \\
  $\omega_1$ & 0.0129 & 0\\
  $\omega_2$ & 0.0127 & 0\\
  $2\omega_1$ & 0.0035 & 0.1097\\
  $2\omega_2$ & 0.0038 & 0.0041\\
  $\omega_1+\omega_2$ & 0.0035 & 0.0469\\
  $4\omega_1$ & $<10^{-4}$ & 0.0233\\
  $3\omega_1+\omega_2$ & $<10^{-4}$ & 0.0200\\ \hline
\end{tabular}
\caption[Table of Franck-Condon factors for tropolone]{Most significant Franck-Condon factors estimated by our experiment and by simulations for an ideal experiment.}
\label{ExpResults}
\end{table}

\subsubsection{Error analysis and classicality}

We apply equation \ref{VSgeneralInequality} to our experiment to determine the theoretical error bound using our analysis of the fidelity and the observed deviation from Gaussian behavior in our system. Given the large number of samples, we neglect the sampling error. We find a bound for the trace distance of 0.455. Our experimental results are indeed within this bound. 

We also apply our classicality criterion to this experiment. The classical state with the highest fidelity to the target state is vacuum, which has a fidelity of 0.879 to the target state. Our experiment has a higher fidelity by about 10 standard deviations and therefore satisfies our classicality criterion. Using vacuum, the classical approximation algorithm described above would yield an error bound of 0.476, which is worse than what we achieved in experiment. However, we note that since vacuum is a Gaussian state, then in the specific case where the closest classical state is vacuum the difference in photon number statistics between the target optical state and the closest classical state can be efficiently calculated, as opposed to simply bounded using the fidelity.

\begin{figure*}[ht]
\includegraphics[width=15cm,angle=0]{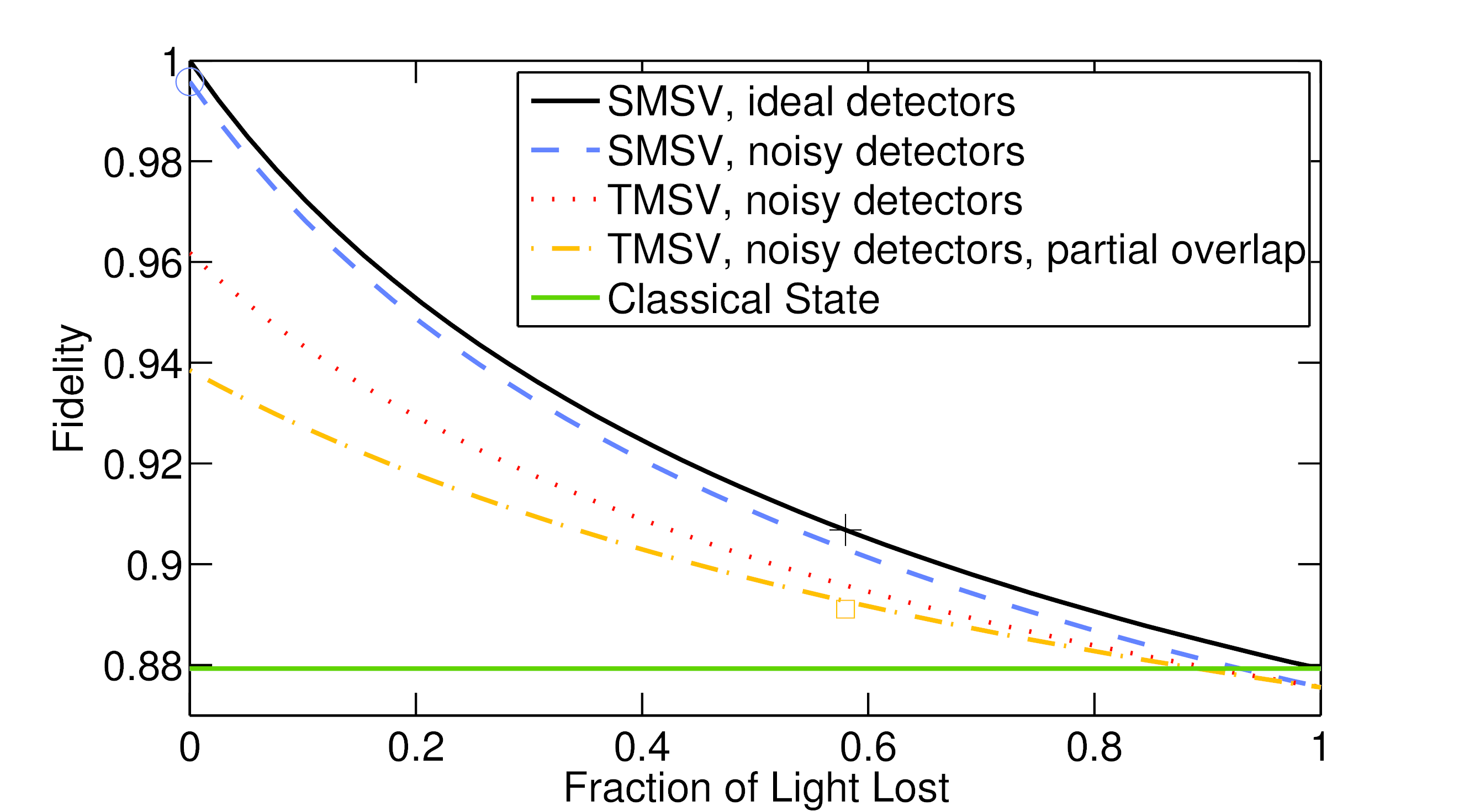}
\caption[Fidelity as a function of loss]{Maximum achievable fidelity as a function of the fraction of light lost in a simulation of tropolone, for different imperfections in an experimental setup. We simulate an ideal setup that is only affected by loss (black solid line), to which we add our noisy detectors (blue dashed line), then replace the SMSVs by a TMSV (red dotted line), then add our measured partial overlap between the modes (orange dashed-dotted line). The cross and the circle respectively indicate the best SMSV for our level of loss and the best SMSV without loss but with noisy detectors, yielding the estimated Franck-Condon factors shown in Table 1. The orange square indicates the experimental parameters used for our experiment. For this analysis, we assume equal loss in both modes, hence the discrepancy between the orange square and the position of the orange dashed-dotted line.}
\label{fig3}
\end{figure*}

\section{Discussion}

In order to contribute to future designs of experiments, in this section we discuss our experimental results, and in particular we analyse the main ways in which different experimental imperfections contribute to the degradation of the fidelity. Figure \ref{fig3} shows the effect of different sources of imperfection on the fidelity in our experimental setup. An ideal experiment consists of SMSVs and ideal detectors (black solid line). The additional dashed and dotted lines indicate the effect of additional imperfections that must be accounted for in our experiment. The orange dashed and dotted line indicates the effect of all the parameters that we account for in our analysis, and the orange square corresponds to our experiment. The flat green solid line indicates the maximum fidelity that can be achieved with the best classical state. These theoretical curves can easily be derived using our Gaussian model for the experiment.

We see that our classicality criterion is relatively tolerant of experimental imperfections to simulate transitions that involve large amounts of squeezing. An enhancement over the best classical state can be achieved with values of loss up to 90\%, with noisy detectors, and with the use of a TMSV instead of two SMSVs. For loss exceeding 90\%, the noise on our detectors degrades the fidelity below that of the best classical state. For transitions in other molecules that involve less squeezing, such as the $\mathrm{S_0 (^{1}A_{1g}) \rightarrow S_1 (^{1}B_{2u})}$ transition in benzene or the $\text{SO}_2^- \rightarrow \text{SO}_2$ transition studied by Shen \textit{et al} \cite{shen2017quantum}, we expect that our level of detector noise would prevent us from outperforming this criterion for any level of loss.

\begin{table}[ht]
\centering
\begin{tabular}{ l c c c r }
  Frequency & \textbf{Experiment} & \textbf{Ideal} & Lossy SMSVs & Best SMSVs  \\ \hline
  0 & 0.9628 & 0.7731 & 0.9327 & 0.7631 \\
  $\omega_1$ & 0.0129 & 0 & 0.0377 & 0.0015 \\
  $\omega_2$ & 0.0127 & 0 & 0.0073 & 0.0015 \\
  $2\omega_1$ & 0.0035 & 0.1097 & 0.0136 & 0.1102 \\
  $2\omega_2$ & 0.0038 & 0.0041 & 0.0004 & 0.0046 \\
  $\omega_1+\omega_2$ & 0.0035 & 0.0469 & 0.0053 & 0.0466 \\
  $4\omega_1$ & $<10^{-4}$ & 0.0233 & 0.0004 & 0.0234 \\
  $3\omega_1+\omega_2$ & $<10^{-4}$ & 0.0200 & 0.0003 & 0.0199 \\ \hline
  Fidelity & 0.890(1) & 1 & 0.9068 & 0.9958 \\ \hline
  Error & 0.206 & 0 & 0.195 & 0.005 \\ \hline
\end{tabular}
\caption[Table of Franck-Condon factors for tropolone]{Most significant Franck-Condon factors, fidelities, and errors estimated by our experiment, and by simulations for an ideal experiment, the best SMSVs for our level of loss, and the best SMSV in a lossless experiment but with noisy detectors.}
\label{FCforTropolone}
\end{table}

A comparison of our experimental results and of the ideal theoretical results for a perfect experiment to simulations of other intermediate states is shown in Table \ref{FCforTropolone}. The loss leads to an overestimate of the Franck-Condon factor corresponding to vacuum, both for our experiment and for a theoretical lossy SMSV. Furthermore, whereas the Franck-Condon factors for odd numbers of excitations should be $0$ due to the difference in symmetry between the ground state and the odd-numbered excited states, our experiment finds these to be non-zero due to photons which would correspond to higher order Franck-Condon factors being lost. The use of a TMSV instead of two independent SMSVs causes us to experimentally find photon numbers that are roughly symmetric in both modes, to within the imbalance in the loss in the two arms. We see from our results that, in the case of tropolone, although the high squeezing and absence of displacement has allowed us to highlight the issue of imperfections in an experiment, by the same token our simulations result in a large error in estimating Franck-Condon factors.

\section{Conclusion}

We have proposed a method for accounting for experimental imperfections in quantum optical simulations of vibronic spectroscopy, following the proposal by Huh \textit{et al}. We have shown that the impact of these imperfections can be quantified, that an experimental setup can be adjusted to account for the presence of these imperfections, and that a classicality benchmark that experiments must outperform to be worthwhile can be formulated. We illustrated these results using a proof of principle experiment that simulated part of the vibronic spectrum of tropolone.

Our results inform future efforts for performing larger scale simulations of vibronic spectra with current quantum optics technology, for example using recent advances in fiber-loop based experiments \cite{inagaki2016coherent,mcmahon2016fully} and integrated photonics \cite{carolan2015universal}. We also note that our approach for dealing with imperfections can be applied to the other platforms which have been proposed for vibronic spectra estimation \cite{hu2017simulating,shen2017quantum}. We envisage that our work will be useful for efficiently approximating the spectra of large molecules that are outside the reach of classical computers.

\section*{Acknowledgements}
We thank David Phillips, David Drahi, Gil Triginer and Johan Fopma for experimental assistance, and Adrian Menssen, Benjamin Metcalf, Peter Humphreys, Raul Garcia-Patron, Jan Sperling, Oscar Dahlsten and Myungshik Kim for helpful discussions. W.R.C, J.J.R, A.E., W.S.K. and I.A.W. acknowledge support from the European Research Council, the UK Engineering and Physical Sciences Research Council (project EP/K034480/1 and the Networked Quantum Information Technology Hub), the Fondation Wiener Anspach, and from the European Commission (H2020-FETPROACT-2014 grant QUCHIP). J.J.R. acknowledges support from NWO Rubicon. J.H. acknowledges support from the Basic Science Research Program through the National Research Foundation of Korea (NRF) funded by the Ministry of Education, Science and Technology (NRF-2015R1A6A3A04059773).

\section*{Additional Information}

\subsection*{Competing financial interests}
The authors declare no competing financial interests.

\subsection*{Supplementary Materials}
Accompanies this paper. The underlying data and models used in this work are available upon reasonable request from the authors.

\bibliographystyle{ieeetr}
\bibliography{referencesVibronicSpectra}

\begin{thebibliography}{10}

\bibitem{aspuru2005simulated}
A.~Aspuru-Guzik, A.~D. Dutoi, P.~J. Love, and M.~Head-Gordon, ``Simulated
  quantum computation of molecular energies,'' {\em Science}, vol.~309,
  no.~5741, pp.~1704--1707, 2005.

\bibitem{huh2015boson}
J.~Huh, G.~G. Guerreschi, B.~Peropadre, J.~R. McClean, and A.~Aspuru-Guzik,
  ``{Boson sampling for molecular vibronic spectra},'' {\em Nature Photonics},
  vol.~9, pp.~615--620, 2015.

\bibitem{huh2017vibronic}
J.~Huh and M.-H. Yung, ``Vibronic boson sampling: Generalized gaussian boson
  sampling for molecular vibronic spectra at finite temperature,'' {\em
  Scientific Reports}, vol.~7, no.~7462, 2017.

\bibitem{hachmann2011harvard}
J.~Hachmann, R.~Olivares-Amaya, S.~Atahan-Evrenk, C.~Amador-Bedolla, R.~S.
  S{\'a}nchez-Carrera, A.~Gold-Parker, L.~Vogt, A.~M. Brockway, and
  A.~Aspuru-Guzik, ``The {H}arvard clean energy project: large-scale
  computational screening and design of organic photovoltaics on the world
  community grid,'' {\em The Journal of Physical Chemistry Letters}, vol.~2,
  no.~17, pp.~2241--2251, 2011.

\bibitem{butler2016using}
H.~J. Butler, L.~Ashton, B.~Bird, G.~Cinque, K.~Curtis, J.~Dorney,
  K.~Esmonde-White, N.~J. Fullwood, B.~Gardner, P.~L. Martin-Hirsch, M.~J.
  Walsh, M.~R. McAinsh, N.~Stone, and F.~L. Martin, ``Using {R}aman
  spectroscopy to characterize biological materials,'' {\em Nature protocols},
  vol.~11, no.~4, pp.~664--687, 2016.

\bibitem{gross2000improving}
M.~Gross, D.~C. M{\"u}ller, H.-G. Nothofer, U.~Scherf, D.~Neher,
  C.~Br{\"a}uchle, and K.~Meerholz, ``Improving the performance of doped
  $\pi$-conjugated polymers for use in organic light-emitting diodes,'' {\em
  Nature}, vol.~405, no.~6787, pp.~661--665, 2000.

\bibitem{janszky1994competition}
J.~Janszky, A.~V. Vinogradov, I.~A. Walmsley, and J.~Mostowski, ``Competition
  between geometrical and dynamical squeezing during a {Franck-Condon}
  transition,'' {\em Physical Review A}, vol.~50, pp.~732--740, Jul 1994.

\bibitem{aaronson2011computational}
S.~Aaronson and A.~Arkhipov, ``The computational complexity of linear optics,''
  in {\em Proceedings of the forty-third annual ACM symposium on Theory of
  computing}, pp.~333--342, ACM, 2011.

\bibitem{spring2012boson}
J.~B. Spring, B.~J. Metcalf, P.~C. Humphreys, W.~S. Kolthammer, X.-M. Jin,
  M.~Barbieri, A.~Datta, N.~Thomas-Peter, N.~K. Langford, D.~Kundys, J.~C.
  Gates, B.~J. Smith, P.~G.~R. Smith, and I.~A. Walmsley, ``Boson sampling on a
  photonic chip,'' {\em Science}, vol.~339, no.~6121, pp.~798--801, 2013.

\bibitem{carolan2015universal}
J.~Carolan, C.~Harrold, C.~Sparrow, E.~Mart{\'\i}n-L{\'o}pez, N.~J. Russell,
  J.~W. Silverstone, P.~J. Shadbolt, N.~Matsuda, M.~Oguma, M.~Itoh, G.~D.
  Marshall, M.~G. Thompson, J.~C.~F. Matthews, T.~Hashimoto, J.~L.
  O{\textquoteright}Brien, and A.~Laing, ``Universal linear optics,'' {\em
  Science}, vol.~349, no.~6249, pp.~711--716, 2015.

\bibitem{wang2017high}
H.~Wang, Y.~He, Y.-H. Li, Z.-E. Su, B.~Li, H.-L. Huang, X.~Ding, M.-C. Chen,
  C.~Liu, J.~Qin, J.-P. Li, Y.-M. He, C.~Schneider, M.~Kamp, C.-Z. Peng,
  S.~Höfling, C.-Y. Lu, and J.-W. Pan, ``High-efficiency multiphoton boson
  sampling,'' {\em Nature Photonics}, vol.~11, no.~6, pp.~361--365, 2017.

\bibitem{hu2017simulating}
L.~Hu, Y.-C. Ma, Y.~Xu, W.-T. Wang, Y.-W. Ma, K.~Liu, H.-Y. Wang, Y.-P. Song,
  M.-H. Yung, and L.-Y. Sun, ``Simulation of molecular spectroscopy with
  circuit quantum electrodynamics,'' {\em Science Bulletin}, 2018.

\bibitem{shen2017quantum}
Y.~Shen, Y.~Lu, K.~Zhang, J.~Zhang, S.~Zhang, J.~Huh, and K.~Kim, ``Quantum
  optical emulation of molecular vibronic spectroscopy using a trapped-ion
  device,'' {\em Chemical Science}, 2018.

\bibitem{wu1986generation}
L.-A. Wu, H.~Kimble, J.~Hall, and H.~Wu, ``Generation of squeezed states by
  parametric down conversion,'' {\em Physical Review Retters}, vol.~57, no.~20,
  p.~2520, 1986.

\bibitem{mcmahon2016fully}
P.~L. McMahon, A.~Marandi, Y.~Haribara, R.~Hamerly, C.~Langrock, S.~Tamate,
  T.~Inagaki, H.~Takesue, S.~Utsunomiya, K.~Aihara, R.~L. Byer, M.~M. Fejer,
  H.~Mabuchi, and Y.~Yamamoto, ``A fully-programmable 100-spin coherent {I}sing
  machine with all-to-all connections,'' {\em Science}, 2016.

\bibitem{inagaki2016coherent}
T.~Inagaki, Y.~Haribara, K.~Igarashi, T.~Sonobe, S.~Tamate, T.~Honjo,
  A.~Marandi, P.~L. McMahon, T.~Umeki, K.~Enbutsu, O.~Tadanaga, H.~Takenouchi,
  K.~Aihara, K.-i. Kawarabayashi, K.~Inoue, S.~Utsunomiya, and H.~Takesue, ``A
  coherent {I}sing machine for 2000-node optimization problems,'' {\em
  Science}, vol.~354, no.~6312, pp.~603--606, 2016.

\bibitem{giddings1976advances}
J.~C. Giddings, {\em Advances in chromatography}, vol.~14.
\newblock CRC Press, 1976.

\bibitem{smith1998strong}
W.~Smith, ``The strong {D}uschinsky effect and the intensity of transitions in
  non-totally symmetric vibrations in the electronic spectra of polyatomic
  molecules,'' {\em Journal of molecular spectroscopy}, vol.~187, no.~1,
  pp.~6--12, 1998.

\bibitem{duschinsky1937importance}
F.~Duschinsky, ``{The importance of the electron spectrum in multi atomic
  molecules. Concerning the Franck-Condon principle},'' {\em Acta Physicochim.
  URSS}, vol.~7, pp.~551--566, 1937.

\bibitem{condon1928nuclear}
E.~U. Condon, ``{Nuclear motions associated with electron transitions in
  diatomic molecules},'' {\em Physical Review}, vol.~32, no.~6, p.~858, 1928.

\bibitem{franck1926}
J.~Franck and E.~G. Dymond, ``{Elementary processes of photochemical
  reactions},'' {\em Trans. Faraday Soc.}, vol.~21, no.~February, pp.~536--542,
  1926.

\bibitem{doktorov1977dynamical}
E.~V. Doktorov, I.~A. Malkin, and V.~I. Man'ko, ``{Dynamical symmetry of
  vibronic transitions in polyatomic molecules and the Franck-Condon
  principle},'' {\em Journal of Molecular Spectroscopy}, vol.~64, no.~2,
  pp.~302--326, 1977.

\bibitem{braunstein2005squeezing}
S.~L. Braunstein, ``{Squeezing as an irreducible resource},'' {\em Physical
  Review A}, vol.~71, no.~5, p.~55801, 2005.

\bibitem{adesso2014continuous}
G.~Adesso, S.~Ragy, and A.~R. Lee, ``Continuous variable quantum information:
  Gaussian states and beyond,'' {\em Open Systems \& Information Dynamics},
  vol.~21, no.~01n02, p.~1440001, 2014.

\bibitem{weedbrook2012gaussian}
C.~Weedbrook, S.~Pirandola, R.~Garc{\'\i}a-Patr{\'o}n, N.~J. Cerf, T.~C. Ralph,
  J.~H. Shapiro, and S.~Lloyd, ``Gaussian quantum information,'' {\em Reviews
  of Modern Physics}, vol.~84, no.~2, p.~621, 2012.

\bibitem{banchi2015quantum}
L.~Banchi, S.~L. Braunstein, and S.~Pirandola, ``Quantum fidelity for arbitrary
  gaussian states,'' {\em Physical Review Letters}, vol.~115, no.~26,
  p.~260501, 2015.

\bibitem{nielsen2010quantum}
M.~A. Nielsen and I.~L. Chuang, {\em Quantum computation and quantum
  information}.
\newblock Cambridge university press, 2010.

\bibitem{laing2012super}
A.~Laing and J.~L. O'Brien, ``Super-stable tomography of any linear optical
  device,'' {\em arXiv preprint arXiv:1208.2868}, 2012.

\bibitem{jankowiak2007}
H.-C. Jankowiak, J.~L. Stuber, and R.~Berger, ``Vibronic transitions in large
  molecular systems: Rigorous prescreening conditions for {F}ranck-{C}ondon
  factors,'' {\em J. Chem. Phys.}, vol.~127, p.~234101, 2007.

\bibitem{leonhardt1997measuring}
U.~Leonhardt, {\em Measuring the quantum state of light}, vol.~22.
\newblock Cambridge university press, 1997.

\bibitem{marian2002quantifying}
P.~Marian, T.~A. Marian, and H.~Scutaru, ``Quantifying nonclassicality of
  one-mode {Gaussian} states of the radiation field,'' {\em Physical Review
  Letters}, vol.~88, no.~15, p.~153601, 2002.

\bibitem{rahimi2016sufficient}
S.~Rahimi-Keshari, T.~C. Ralph, and C.~M. Caves, ``Sufficient conditions for
  efficient classical simulation of quantum optics,'' {\em Physical Review X},
  vol.~6, no.~2, p.~021039, 2016.

\bibitem{rahimi2015can}
S.~Rahimi-Keshari, A.~P. Lund, and T.~C. Ralph, ``What can quantum optics say
  about computational complexity theory?,'' {\em Physical Review Letterrs 114,
  060501 (2015)}, 2015.

\bibitem{eckstein2011highly}
A.~Eckstein, A.~Christ, P.~J. Mosley, and C.~Silberhorn, ``Highly efficient
  single-pass source of pulsed single-mode twin beams of light,'' {\em Physical
  Review Letters}, vol.~106, no.~1, p.~013603, 2011.

\bibitem{lita2008counting}
A.~E. Lita, A.~J. Miller, and S.~W. Nam, ``Counting near-infrared
  single-photons with 95\% efficiency,'' {\em Optics express}, vol.~16, no.~5,
  pp.~3032--3040, 2008.

\bibitem{hong1987measurement}
C.~K. Hong, Z.-Y. Ou, and L.~Mandel, ``Measurement of subpicosecond time
  intervals between two photons by interference,'' {\em Physical review
  letters}, vol.~59, no.~18, p.~2044, 1987.

\bibitem{worsley2009absolute}
A.~Worsley, H.~Coldenstrodt-Ronge, J.~Lundeen, P.~Mosley, B.~Smith, G.~Puentes,
  N.~Thomas-Peter, and I.~Walmsley, ``Absolute efficiency estimation of
  photon-number-resolving detectors using twin beams,'' {\em Optics express},
  vol.~17, no.~6, pp.~4397--4412, 2009.

\end{thebibliography}

\end{document}